%% file: main.tex
\documentclass{article}
\usepackage{authblk}
\usepackage[utf8]{inputenc}
\usepackage{latexsym}
\usepackage[english]{babel}
\usepackage[margin=1.0in]{geometry}
\usepackage{float}
\floatstyle{boxed} 
\usepackage{amsmath}
\usepackage{amsfonts}
\usepackage{amssymb}
\usepackage{dsfont}
\usepackage{tikz}
\usetikzlibrary{decorations.pathmorphing}
\usepackage{pgfplots}
\usepackage{caption}
\usepackage{standalone}
\usepackage[superscript, biblabel, nomove]{cite}
\begin{document}
\title{The Zubarev Double Time Green's function\\-A Vintage Many Body Technique}
\author[1]{Vijay Singh\thanks{$^\ast$physics.sutra@gmail.com}}
\author[2]{Shraddha Singh}
\affil[1]{Department of Physics, UM-DAE, Centre for Excellence in Basic Sciences, Mumbai University, Kalina, Santa-Cruz East, Mumbai 400098, India.}
\date{}
\maketitle
\begin{abstract}
These lecture notes present a comprehensive and powerful many-body technique pioneered in 1960 by D. N. Zubarev. The technique, known as the Zubarev Double Time Green's Function method, was used extensively by leading solid state physicists such as John Hubbard and Laura Roth in the 1960s.  We present the technique and apply it to the non-interacting electron and boson gas. We next consider the (many-body) Hubbard model and show how it yields the Stoner criterion for ferromagnetism. It is easily extendable to superconductivity and related problems.  Our treatment is pedagogical and understandable to those with just an  elementary understanding of second quantization. 
\end{abstract}
\tableofcontents
%\listoffigures
\section{Introduction}
 The development of Quantum Many Body Techniques employing second quantized formalism had their early beginnings in the 1950s. These were inspired  by, but by no means clones of the perturbative and diagrammatic approach in quantum field theory. Pioneering approaches in  any field tend to be tentative, exploratory and even error prone. So, Zubarev's path breaking paper\cite{11}, with its  completeness, elegance and sparse character, appearing in 1960, comes as a big surprise. Indeed his Double Time Green's formalism leaves little for others to do. It captured the imagination of condensed matter theorists both in the east and in the west. Perhaps, an example may not be out of place here.
\par A model interaction Hamiltonian which is arguably the most famous of all is the Hubbard Hamiltonian. Proposed in 1963, only a few would know nowadays that Hubbard employed Zubarev's Double Time Green's function to obtain approximate solutions to his model, famously known as Hubbard I\cite{21} and Hubbard III. \cite{31}. In this context, a mention must also be made of Roth's\cite{41} innovative approach to the same Hamiltonian which also made use of Zubarev's formulation.  

%Definition and Notation
\subsection {Definition and Notation}
Zubarev's Retarded Green's function is defined as follows-
\begin{eqnarray}
G^{r}_{AB}(t,t')=\langle\langle A(t);B(t')\rangle\rangle^{r}=-i\theta(t-t')\langle[A(t),B(t')]_{\eta}\rangle,
\label{a}
\end{eqnarray}
where A(t), B(t') are operators with 
$$A(t)=exp(-iHt)A(0)exp(iHt)$$
 and likewise for B(t') with the Hamiltonian H being independent of time. Note that we have taken $\hbar$ = 1.
\begin{eqnarray*}
[A(t),B(t')]_{\eta}&=&A(t)B(t')-\eta B(t')A(t'),\text{ where\footnotemark}\label{l}
\end{eqnarray*}
\footnotetext{To clear a possible confusion here, we would like to bring to the readers' notice that while using this formula one must not substitute the value of $\eta$ in L.H.S. but only use it in R.H.S. for the corresponding L.H.S. Short hand notation for a given $\eta$ could be used from Eqs.(\ref {m1}) and (\ref{m2}). The confusion arose due to the $\eta$ formalism used in Zubarev's paper. We shall continue to use it whenever, unavoidable. Note, textbooks use the standard formalism given by Eqs.(\ref{m1}) and (\ref{m2}) and we shall adopt it whenever possible.}
\begin{subequations}
\begin{eqnarray}
 \eta &=&+1,\text{     for boson operators}\label{l1}\\
&=&-1,\text{     for fermion operators}\label{l2}
\end{eqnarray}
\end{subequations}
Note that we may also use
\begin{subequations} 
\begin{eqnarray}
[A(t),B(t')]_{-}&=&A(t)B(t')-B(t')A(t),\text{    for commutator}\label{m1}\\ 
\quad[A(t),B(t')]_{+}&=&A(t)B(t')+B(t')A(t),\text{    for anti commutator}\label{m2}
\end{eqnarray}
\end{subequations}
Also, in the canonical representation  which is the case under consideration in this article, $\langle A(t)\rangle=Z^{-1} tr\{\mbox{A(t)exp}(-\beta H)\}$. In the grand canonical representation, $\beta H\longrightarrow\beta (H-\mu N)$, $\mu$ being the chemical potential. Further, as mentioned earlier, we take $\hbar=1$. Note, $tr$ stands for the trace and $\beta = 1/k_{B}T$, where $k_{B}$ is Boltzmann constant. 
\par The symbol $\theta$(t) denotes the step function and is s illustrated below in Fig.(\ref{b7}).
\begin{eqnarray*}
\theta(t)&=&1, \quad\quad t>0 \\
&=&0, \quad\quad t<0
\end{eqnarray*}
We note that,
\begin{eqnarray}
\dfrac{d}{dt}\theta(t)=\delta(t),
\label{b}
\end{eqnarray}
 where $\delta$(t) is the Dirac delta function.
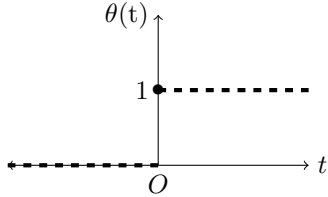
\begin{figure}[h]
\centering
\input{tikz11}
\caption{The dashed line represents the step function $\theta$(t)}
\label{b7}
\end{figure}
%Physical Picture
\subsection{Physical Picture}\label{s5}
In this section we help the reader to develop a  conceptual  understanding of the present work. To a beginner, the Green's function in Eq.(\ref{a}) would look unfamiliar. Rather, our experience with electrostatics  would lead us to expect an inverse operator. In electrostatics the Green's function is an ``\textit{inverse Laplacian}''. It is the time Fourier transform of the Green's function defined above that resembles this inverse. In passing, we mention that the Green's function is variously referred to as the kernel,   correlation function, resolvent, propagator or locator and each of these terms has a special significance and specific meaning. Some of these will become clear as we progress. 

\begin{figure}[h]
\centering
\input{tikz21}
\caption{As the electron "traverses'' the system its spin and momentum may both undergo changes. The Green's function acts as a probe. }
\label{b8}
\end{figure}
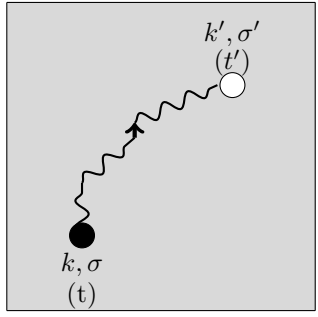

A legitimate question at this stage would be to ask what the operators A(t) and B(t') are? In condensed matter theory, possible candidates are  electron annihilation ($c_{i\sigma}$) and electron creation ($c^{\dagger}_{i\sigma}$) operators. An example is
\begin{eqnarray*}
G^{r}_{kk',\sigma\sigma '}(t,t')=\langle\langle c_{k\sigma}(t);c^{\dagger}_{k'\sigma'}(t')\rangle\rangle^{r}
\end{eqnarray*}
This describes the creation of an electron with momentum k$'$, spin $\sigma'$, at time t$'$ and its subsequent annihilation at time t, with spin $\sigma$  and momentum k. The spins and the momenta can both undergo change as the electron 'traverses'' the system. What is meant by "the system''? The system is represented by the Hamiltonian. Thus the Green's function acts as a probe of the system. The traversal yields useful information on elementary excitation of the system. Fig.(\ref{b8}) depicts this pictorially.

\par The system as mentioned above is represented by the Hamiltonian. In the beginning we mentioned a $"$famous$''$ model Hamiltonian. namely the Hubbard Hamiltonian. Here is how it looks:
\begin{eqnarray*}
H=\mu\sum_{i\sigma}n_{i\sigma}-t\sum_{i\neq j \sigma}^{n.n}c^{\dagger}_{i\sigma}c_{j\sigma}+\frac{U}{2}\sum_{i\sigma}n_{i\sigma}n_{i-\sigma},
\end{eqnarray*}
Here $\sigma$ represents the spin of the particles. It can take two values, spin up $\uparrow(+)$ and spin down $\downarrow(-)$. The number operator is $n_{i\sigma} = c^{\dagger}_{i\sigma}c_{i\sigma}$.  The first term represents the on - site energy of the electron, where $\mu$ is also called the chemical potential. Think of a one-dimensional row of hydrogen atoms. Then $\mu$ is like our ground state energy, -13.6 eV. The second term represents hopping from site 'i' to site 'j'. The symbol 't' stands for transfer. In the Hubbard model we allow only nearest neighbour transfer - the electron can hop from site 'i'' to 'i+1' or 'i-1' and not to 'i+2' or 'i-3' etc. The term  U represents the Coulomb repulsion  which prevents an electron from hopping to an already occupied  neighboring site. Note that only 1s-orbitals are present so a given site can,  as per Pauli exclusion principle, be occupied by at most 2 electrons. Also we are used to thinking of the Coulomb interaction as long range, but in this case it is highly localized and does not extend to even nearest neighbour! In soild state parlance we say that the Coulomb interaction is screened. We have simplified the discussion here and if one is interested a better place to look at is  the introductory sections of Hubbard's first paper\cite{21}. 

 If we set $U = 0$ in the above Hamiltonian then you have what is called as a non-interacting system or equivalently a free electron gas system. Examples of the free electron gas and phonon gas will be taken up in in Sec.\ref{s1} and this would clarify the meaning of Zubarev Green's function further.   

\section{Equation of motion}
Let's see how one may obtain the Green's function. Note $\hbar=1$ and $A(t)=exp(-iHt)A(0)exp(iHt)$,
\begin{eqnarray}
i\frac{d}{dt}A=[A(t),H]_{-},
\label{a0}
\end{eqnarray}
Using Eqs.(\ref{a}) and (\ref{b}), we obtain
\begin{eqnarray*}
i\frac{d}{dt}G(t,t')&=&\delta(t-t')\langle[A(t),B(t')]_{\eta}\rangle -i \theta (t-t')\langle[A(t),H]_{-}B(t')-\eta B(t')[A(t),H]_{-}\rangle 
\end{eqnarray*}
\begin{eqnarray}
&=&\delta(t-t')\langle[A(t),B(t')]_{\eta}\rangle+\langle\langle[A(t),H]_{-};B(t')\rangle\rangle^{r}
\label{i}
\end{eqnarray}
resulting in a higher order Green's function. If one repeats the same step then,
\begin{equation*}
i\frac{d}{dt}\langle\langle[A(t),H]_{-};B(t')\rangle\rangle^{r}=\delta(t-t')\langle[[A(t),H]_{-},B(t')]_{\eta}\rangle+ \langle\langle[[A(t),H]_{-},H]_{-};B(t')\rangle\rangle^{r}
\end{equation*}
\par This exercise  yields a hierarchy of higher order Green's functions. Unless the Hamiltonian is simple (see Sec.\ref{s1}), this exercise can go on and on. In general we cut this Gordian knot by imposing on physical grounds or otherwise the condition, $\langle\langle[[A(t),H]_{-},H]_{-};B(t')\rangle\rangle^{r}\approx\alpha\langle\langle[A(t),H]_{-};B(t')\rangle\rangle^{r}$, where $\alpha$ is a scalar. This exercise is called \textit{truncation} or \textit{decoupling}. It results in a closed set of equations.  

In hindsight we should not be surprised. The many body problem is essentially intractable and this is reflected in this infinitely recursive situation of the Zubarev technique. In related many body techniques it shows up in the form of an infinite series or an infinite set of Feynman diagrams. Before we take up cases of how we actually deal with this, we need to develop the formalism further.

\section{The Formalism}
%Correlation Functions

\subsection{Correlation Function}
The Green's function is related to the correlation functions as follows;
\begin{eqnarray*}
F_{AB}(t-t')=\langle A(t)B(t')\rangle \\
F_{BA}(t-t')=\langle B(t')A(t)\rangle
\end{eqnarray*}
\begin{eqnarray*}
G^{r}_{AB}(t,t')=-i\theta(t-t')\{F_{AB}(t-t')-\eta F_{BA}(t-t')\}
\end{eqnarray*} 
Henceforth, we will assume a $"$(t-t')$''$ dependence (justified later in Sec. \ref{s9}). The terminology, namely correlation functions should not surprise us. For example, we use spin-spin correlation function for the  Ising model as $\langle S_iS_j\rangle$, where \{$S_i$, $S_j$\} are classical variables. It yields information about the magnetic nature of the system. In our case  \{A(t), B(t')\} are quantum operators.

\subsection{Spectral Representation of the Green's Function}\label{s9}
The Green's function was defined in terms of the \textit{trace}. It is customary to express this \textit{trace} in terms of the eigen functions $\{|\phi_{n}\rangle\}$ of the Hamiltonian. This exercise is called the spectral decomposition and we carry it out presently. We begin with the Schrodinger equation,
\begin{eqnarray*}
H\phi_{n}=E\phi_{n},
\end{eqnarray*}
Taking the expectation value of the the operators occurring in the correlation function, we get the following equations,
\begin{eqnarray*}
F_{BA}(t-t')&=&Z^{-1} tr (B(t')A(t)e^{-\beta H})\\
&=&Z^{-1}\sum_{n}\langle\phi_{n}|B(t') A(t)e^{-\beta H}|\phi_{n}\rangle\\
\end{eqnarray*}

Next, we introduce a complete set of eigen states to obtain,
\begin{eqnarray*}
F_{BA}(t-t')&=&Z^{-1}\sum_{n,m}\langle\phi_{n}|B(t')|\phi_{m}\rangle\langle\phi_{m}|A(t)e^{-\beta H}|\phi_{n}\rangle\\
&=&Z^{-1}\sum_{n,m}\langle\phi_{n}|e^{iHt'}B(0)e^{-iHt'}|\phi_{m}\rangle\langle\phi_{m}|e^{iHt}A(0)e^{-iHt}|\phi_{n}\rangle e^{-\beta E_{n}}\\
&=&Z^{-1}\sum_{n,m}e^{i(E_n-E_m)t'}B_{nm}e^{i(E_m-E_n)t}A_{mn}e^{-\beta E_{n}}
\end{eqnarray*}
Thus,
\begin{eqnarray}
F_{BA}(t-t')&=&Z^{-1}\sum_{n,m}e^{-i(E_n-E_m)(t-t')}B_{nm}A_{mn}e^{-\beta E_{n}},
\label{c}
\end{eqnarray}
where,
\begin{eqnarray*}
B_{nm}&=&\langle\phi_{n}|B(0)|\phi_{m}\rangle\\
A_{mn}&=&\langle\phi_{m}|A(0)|\phi_{n}\rangle
\end{eqnarray*}

Similarly,
\begin{eqnarray}
F_{AB}(t-t')=Z^{-1}\sum_{n,m}e^{i(E_n-E_m)(t-t')}B_{mn}A_{nm}e^{-\beta E_{n}}
\label{d}
\end{eqnarray}

Note that R.H.S. in Eqs.(\ref{c}) and (\ref{d}) depends only on (t-t'), hence it is justified to write $G^{r}_{AB}(t,t')=G^{r}_{AB}(t-t')$

\subsection{Spectral Density}

It is useful to define a spectral weight or spectral density function J($\omega$) as follows
\begin{eqnarray*}
J(\omega)&=&Z^{-1}\sum_{n,m}A_{nm}B_{mn}e^{-\beta E_{m}}\delta(E_m-E_n-\omega)
\end{eqnarray*}
Note that we have taken $\hbar=1$, hence $\hbar\omega=\omega$. This implies,
\begin{eqnarray}
F_{BA}(t-t')&=&\int_{-\infty}^{\infty}J(\omega)e^{-i\omega(t-t')}d\omega\label{k}
\end{eqnarray}
\begin{eqnarray}
F_{AB}(t-t')&=&\int_{-\infty}^{\infty}J(\omega)e^{\beta\omega}e^{-i\omega(t-t')}d\omega\label{z94}
\end{eqnarray}
It maybe useful to verify the spectral representation for one of the expressions say, for $F_{AB}$ 
\begin{eqnarray*}
F_{AB}(t-t')&=&\int_{-\infty}^{\infty}J(\omega)e^{\beta\omega}e^{-i\omega(t-t')}d\omega\\
&=&\int_{\infty}^{-\infty}(Z^{-1}\sum_{n,m}A_{nm}B_{mn}e^{-\beta E_{m}})\delta(E_m-E_n-\omega)e^{\beta\omega}e^{-i\omega(t-t')}d\omega\\
&=&Z^{-1}\sum_{n,m}A_{nm}B_{mn}e^{-\beta E_{m}})e^{\beta (E_m-E_n)}e^{-i(E_m-E_n)(t-t')}\\
&=&Z^{-1}\sum_{n,m}A_{nm}B_{mn}e^{\beta (E_m-E_n-E_m)}e^{-i(E_m-E_n)(t-t')}
\end{eqnarray*}
Therefore,
\begin{eqnarray}
F_{AB}(t-t')&=&Z^{-1}\sum_{n,m}A_{nm}B_{mn}e^{-\beta E_n}e^{-i(E_m-E_n)(t-t')}
\label{e}
\end{eqnarray}
Eq.(\ref{e}) is the same as Eq.(\ref{d}). Hence, we have verified the spectral representation of correlation functions.
%Fourier Transform
\subsection{Relation between Green's Function and The Spectral Density} 
The Green's function in time domain is not commonly found in many-body physics. One does encounter it in say the Feynman and Hibbs book\cite{51} on Feynman integrals where it is called the kernel or propagator and denotes  the probability of a quantum particle traversing from $\{x',t'\}$ to $\{x,t \}$. It is the Fourier transform of the Green's function to the frequency (or energy) domain that is related to observables. We define this as follows,
\begin{eqnarray*} 
G^{r}_{AB}(t,t')=\int_{-\infty}^{\infty}G^{r}_{AB}(E)e^{-iE(t-t')}dE
\end{eqnarray*}
Consequently, the inverse Fourier transform is
\begin{eqnarray*}
G^{r}_{AB}(E)&=&\frac{1}{2\pi}\int_{-\infty}^{+\infty}G^{r}_{AB}(t-t')e^{iE(t-t')}d(t-t')\\
&=&\frac{1}{2\pi}\int_{-\infty}^{\infty}-i\theta(t-t')\{F_{AB}(t-t')-\eta F_{BA}(t-t')\}e^{iE(t-t')}d(t-t')\\\
&=&\frac{-i}{2\pi}\int_{-\infty}^{\infty}\theta(t-t')e^{iE(t-t')}\int_{-\infty}^{\infty}J(\omega)(e^{\beta\omega}-\eta)e^{-i\omega(t-t')}d\omega d(t-t')
\end{eqnarray*}
The last equation is obtained on expressing the correlation functions in terms of the spectral density (see Eqs. (\ref{k}) and (\ref{z94})). Interchanging the $\omega$ and (t-t') integrations, we get

\begin{eqnarray*}
&=&\frac{-i}{2\pi}\int_{-\infty}^{\infty}J(\omega)(e^{\beta\omega}-\eta)d\omega\int_{-\infty}^{\infty}\theta(t-t')e^{i(E-\omega)(t-t')} d(t-t')
\end{eqnarray*}
Substituting integral representation of the step function $\theta(t)$ [Appendix \ref{s2}] we write,
\begin{eqnarray*}
\theta(t)&=&\text{lim}_{\epsilon\rightarrow 0^+}\frac{i}{2\pi}\int_{-\infty}^{\infty}\frac{dx}{x+i\epsilon}e^{-ixt}, \\
G^r(E)&=&\text{lim}_{\epsilon\rightarrow 0^+}\frac{-i}{2\pi}\int_{-\infty}^{\infty}d\omega J(\omega)(e^{\beta\omega}-\eta)\frac{i}{2\pi}\int_{-\infty}^{\infty}\frac{dx}{x+i\epsilon}\int_{-\infty}^{\infty}e^{i(E-\omega-x)(t-t')} d(t-t')\\
&=&\text{lim}_{\epsilon\rightarrow 0^+}\frac{1}{2\pi}\int_{-\infty}^{\infty}d\omega J(\omega)(e^{\beta\omega}-\eta)\int_{-\infty}^{\infty}dx\frac{\delta(E-\omega-x)}{x+i\epsilon}\\
&=&\text{lim}_{\epsilon\rightarrow 0^+}\frac{1}{2\pi}\int_{-\infty}^{\infty}d\omega J(\omega)\frac{(e^{\beta\omega}-\eta)}{E-\omega+i\epsilon}
\end{eqnarray*}
where in the above expression we have used the integral representation of the Dirac $\delta$-function,
\begin{eqnarray}
\delta(\omega) &=& \frac{1}{2\pi}\int_{-\infty}^{\infty} e^{ i\omega t} dt \label{delta}
\end{eqnarray}
\begin{eqnarray}
G^r(E)&=&\frac{1}{2\pi}\int_{-\infty}^{\infty} J(\omega)\frac{(e^{\beta\omega}-\eta)}{E-\omega}d\omega
\label{f}
\end{eqnarray}
Interchanging E and $\omega$ in Eq.(\ref{f}), we have 
\begin{eqnarray*}
G^r(\omega)&=&\frac{1}{2\pi}\int_{-\infty}^{\infty} J(E)\frac{(e^{\beta E}-\eta)}{\omega-E}dE\\
G^r(\omega+i\epsilon)-G^r(\omega-i\epsilon)&=&\frac{1}{2\pi}\int_{-\infty}^{\infty} J(E)(e^{\beta E}-\eta)(-2i\pi\delta(\omega-E))dE,\\
&=&-iJ(\omega)(e^{\beta\omega}-\eta)
\end{eqnarray*}
where the last step is obtained on using the Plemjl formula (Appendix \ref{s3}).
\begin{eqnarray}
 \dfrac{1}{x \pm i\epsilon} = \mbox{Pr }\dfrac{1}{x} \mp i \pi \delta(x) \label{plemjl}
\end{eqnarray}
Thus 
\begin{eqnarray}
J(\omega)&=&i\frac{\{G^r(\omega+i\epsilon)-G^r(\omega-i\epsilon)\}}{e^{\beta\omega}-\eta}
\label{g}\\
&=&\frac{-2ImG^r(\omega)}{e^{\beta\omega}-\eta},
\label{h}
\end{eqnarray}
where we take advantage of the (unproved) analytic property of the Green's function $G^r(\omega-i\epsilon) = G^{r*}(\omega+i\epsilon)$. The imaginary part of the Green's function in the energy domain is related to the spectral density in a simple way. The latter will be shown to be connected in a simple fashion to observables such as the density of states.
 
\subsection{Revisiting the Equation of Motion}
We will rewrite the equation of motion in the energy domain. Fourier transforming the left hand side of  Eq.(\ref{i}) and integrating by parts
\begin{eqnarray*}
\text{L.H.S.}&:&\frac{1}{2\pi}\int_{-\infty}^{\infty}i\frac{d}{dt}G^r(t-t')e^{i\omega(t-t')}d(t-t')\\
&=&\frac{i}{2\pi}G^r(t-t')e^{i\omega(t-t')}\bigg|_{-\infty}^{\infty}+\frac{\omega}{2\pi}\int_{-\infty}^{\infty}G^r(t-t')e^{i\omega(t-t')}d(t-t')
\end{eqnarray*}
The second term is simply $\omega G^r(\omega)$.
The first terms are  boundary terms. They  vanish at both limits. As  $t\rightarrow +\infty$, the upper limit vanishes since      $\omega \rightarrow\omega+i\epsilon$ $(t>0)$. As $t\rightarrow -\infty$, the lower limit vanishes since $\theta(t-t')\Longrightarrow(t>t')$ and the function does not exist. Thus, L.H.S. is  $= \omega G^r(\omega)$. R.H.S. is simpler to deal with,
\begin{eqnarray*}
\omega G^r(\omega)&=&\frac{1}{2\pi}\int_{-\infty}^{\infty}\delta(t-t')e^{i\omega(t-t')}\langle[A,B]_\eta\rangle d(t-t')+\langle\langle[A,H]_{-};B\rangle\rangle^r_\omega
\end{eqnarray*}
\begin{eqnarray}
&=&\frac{1}{2\pi}\langle[A,B]_\eta\rangle+\langle\langle[A,H]_{-};B\rangle\rangle_\omega^r
\label{j}
\end{eqnarray}
The first term is an equal time commutator. Hence, we have obtained the equation of motion (Eq.(\ref{j})) in the energy domain. %Also, from Eq.(\ref{k}) we have
%\begin{eqnarray*}
%\langle B(t'),A(t)\rangle &=&\text{i }\text{lim}_{\epsilon\rightarrow0^+}\int_{-\infty}^\infty\frac{G^r(\omega+i\epsilon)-G^r(\omega-i\epsilon)}{e^{\beta \omega}-\eta}e^{-i\omega(t-t')}d\omega\\
%&=&\text{i }\text{lim}_{\epsilon\rightarrow0^+}\int_{-\infty}^\infty\frac{\langle\langle A,B\rangle\rangle^r_{\omega+i\epsilon}-\langle\langle A,B\rangle\rangle^r_{\omega-i\epsilon}}{e^{\beta \omega}-\eta}e^{-i\omega(t-t')}d\omega
%\end{eqnarray*}
%This is the second equation of motion.

%Examples
\section{Examples}\label{s1}
In the light of Zubarev's double time Green's function method, we solve 'Free Electron Gas' and 'Phonon Gas' Hamiltonians to get the respective distribution functions. This section repeatedly uses some commutation relations stated in Appendix \ref{s4}.
%Free Electron Gas
\subsection{Free Electron Gas- The Perfect Quantum gas}
\begin{figure}[h]
\centering
\input{tikz31}
\caption{ Band Structure }
\label{b9}
\end{figure}
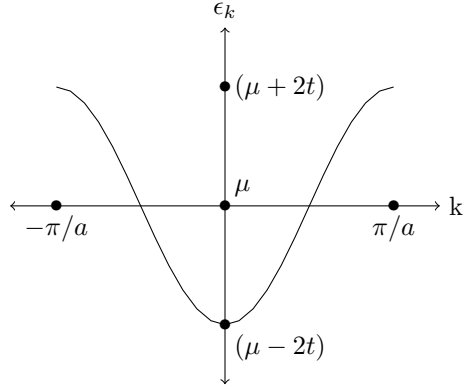

Free Electron model represents a model where electrons do not interact with each other. Thus,  
\begin{eqnarray*}
H=\sum_{k\sigma}\epsilon_k c_{k\sigma}^\dagger c_{k\sigma}
\end{eqnarray*}
Here the system is represented  by $\epsilon_k$, the spin-independent band structure. For a one-dimensional nearest neighbour tight binding model, $\epsilon_k=\mu - 2t cos(ka)$ and it is illustrated in Fig.(\ref{b9}) 
\\To repeat, this Hamiltonian corresponds to a non-interacting system (U=0) as compared to the Hamiltonian described in the physical picture in Sec.(\ref{s5}). Noting that for fermions, $\eta=-1$, we define Green's function for this case as follows.
\begin{eqnarray}
 G^r_{kk'\sigma\sigma'}(t-t')&=&\langle\langle c_{k\sigma}(t),c_{k'\sigma'}(t')^\dagger\rangle\rangle^r \label{n}\\
&=&-i\theta(t-t')\langle[c_{k\sigma}(t); c_{k'\sigma'}(t')^\dagger]_+\rangle,\text{ as }([c_{k\sigma},c_{k'\sigma'}^\dagger]_\eta= [c_{k\sigma},c_{k'\sigma'}^\dagger]_+\text{ from Eqs.(2) and Eq.(3)})\nonumber
\end{eqnarray} 
\begin{eqnarray*}
i\frac{d}{dt}c_{k\sigma}(t)=[c_{k\sigma}(t),H]_-=\epsilon_kc_{k\sigma}(t),\quad\quad\text{(from Eqs.(\ref{a0}), (\ref{a2}) and (\ref{a1})) }
\end{eqnarray*}
Equation of motion is given by the following equation.
\begin{eqnarray*}
i\frac{d}{dt}G^r_{kk'\sigma\sigma'}(t-t')&=&\delta(t-t')\langle[c_{k\sigma},c_{k'\sigma'}^\dagger]_+\rangle+\langle\langle i\frac{d}{dt}c_{k\sigma}(t);c_{k'\sigma'}(t')^\dagger\rangle\rangle^r\\
&=&\delta(t-t')\delta_{\sigma\sigma'}\delta_{kk'}+\epsilon_k\langle\langle c_{k\sigma}(t);c_{k'\sigma'}(t')^\dagger\rangle\rangle^r
\end{eqnarray*}
These equations do not yield higher order Green's functions and hence, approximations are not required. In other words, this is an exact solution. Fourier transforming the above equation we have from Eq.(\ref{j}),
\begin{eqnarray}
\omega G^r_{kk'\sigma\sigma'}(\omega)&=&\frac{1}{2\pi}\delta_{\sigma\sigma'}\delta_{kk'}+\epsilon_k G^r_{kk'\sigma\sigma'}(\omega)\nonumber\\
\therefore \quad G^r_{kk'\sigma\sigma'}&=&\frac{1}{2\pi}\frac{\delta_{\sigma\sigma'}\delta_{kk'}}{\omega-\epsilon_k}\label{a5}
\end{eqnarray}
$c_{k\sigma}^\dagger c_{k\sigma}=n_{k\sigma}$ is the number operator and $\langle n_{k\sigma}\rangle$, the distribution function. Hence, the important information to extract from the derivations carried out in the preceding sections is the correlation function, $\langle c_{k\sigma}^\dagger c_{k'\sigma'}\rangle$. 
\begin{eqnarray*}
\langle c_{k\sigma}^\dagger c_{k'\sigma'}\rangle&=&-2\int_{-\infty}^{\infty}d\omega\frac{\text{Im}G_{kk'\sigma\sigma'(\omega)}}{e^{\beta\omega}+1}\\
&=&-2\int_{-\infty}^{\infty}d\omega\frac{\text{Im}\dfrac{\delta_{kk'}\delta_{\sigma\sigma'}}{2\pi(\omega-\epsilon_k+i\epsilon)}}{e^{\beta\omega}+1}\\
&=&2\pi\int_{-\infty}^{\infty}d\omega \delta(\omega-\epsilon_k)\frac{\delta_{kk'}\delta_{\sigma\sigma'}}{2\pi}\frac{1}{e^{\beta\omega}+1}\quad \quad \text{(using Plemjl formula (See Appendix \ref{s3}))}\\
\therefore \langle c_{k\sigma}^\dagger c_{k'\sigma'}\rangle&=& \frac{\delta_{kk'}\delta_{\sigma\sigma'}}{e^{\beta\epsilon_k}+1}
\end{eqnarray*}

\begin{eqnarray}
\langle n_{k\sigma}\rangle=\frac{1}{e^{\beta\epsilon_k}+1}\label{a6}
\end{eqnarray}
This is the standard Fermi Dirac distribution function.
\newline
\\\textbf{Remarks$\colon$}
\\1.) The choice of operators which go into the Green's function must be judicious. Our choice in Eq.(\ref{n}) was such that the correlation function turned out to be the distribution function $\langle n_{k\sigma}\rangle$.
\\2.) In general, for an interacting or complex system we expect 
\begin{eqnarray*}
G^r_{kk'\sigma\sigma'}=\frac{1}{2\pi(\omega-\epsilon_k-\sum_{k\sigma}(\omega))},
\end{eqnarray*}
where$ \sum_{k\sigma}(\omega)$ is the self energy. Here the energy of elementary excitation is $\omega-\epsilon_k-Re(\sum_{k\sigma}(\omega)))$ and the lifetime is $\hbar/\text{Im}(\sum_{k\sigma}(\omega))$
\\3.) Note that the Green's function is an inverse function. It is first encountered in electrostatics where it is used to obtain the electrostatic potential due to a point charge. This is also true for Zubarev's double time Green's function. Although not apparent in Eq.(\ref{a}) it becomes clear in the fourier transform version in Eq.(\ref{a5}) 
\\4.) Even without the Zubarev technique we could have derived Eq.(\ref{a5}) in a straight forward way by employing the definition of Green's function as an inverse operator,
\begin{eqnarray*}
(\omega\hat I-\hat H)\hat G &=& \dfrac{\hat I}{2\pi}\\
\hat G &=& \dfrac{\hat I}{2\pi}(\omega\hat I-\hat H)^{-1}\\
\hat G &=& \dfrac{\hat I}{2\pi(\omega\hat I-\hat H)}\\
\end{eqnarray*}
Taking the expectation value between momentum states and using $\langle k\sigma|\hat H|k\sigma\rangle = \epsilon_k$, we obtain the Green's function as follows, 
\begin{eqnarray*}
\langle k\sigma|\hat G|k\sigma\rangle &=& \dfrac{1}{2\pi(\omega-\epsilon_k)}
\end{eqnarray*}
%Phonon Gas
\subsection{Phonon Gas}
We discuss phonon gas using the same approach as free electron gas. In this case the Hamiltonian is, 
\begin{eqnarray*}
H=\sum_q(\text{a(q)}^\dagger a(q)+1/2)\hbar\omega_q, 
\end{eqnarray*}
where q represents the wave vector space and a(q)$^\dagger$ (a(q)) are the creation (annihilation) operator. For phonons, $[a(q),a(q')^\dagger]_-=\delta_{qq'}$ and $\eta=1$.
In this case the Green's function is defined as follows.
\begin{eqnarray*}
 G^r_{qq'}&=&\langle\langle a(q); a(q')^\dagger\rangle\rangle^r
\end{eqnarray*}
From the commutation relation for phonon operators and the given Hamiltonian it follows that
\begin{eqnarray*}
[a(q),H]_-=(\hbar\omega_q)a(q) \text{ (using Eq.(\ref{a3}))}
\end{eqnarray*}
Hence, using the equation of motion we get,
\begin{eqnarray*}
\omega\langle\langle a(q); a(q')^\dagger\rangle\rangle^r&=&\frac{\delta_{qq'}}{2\pi}+(\hbar\omega_q)\langle\langle a(q); a(q')^\dagger\rangle\rangle^r\\
G^r_{qq'}&=&\frac{\delta_{qq'}}{2\pi(\omega-\hbar\omega_q)}
\end{eqnarray*}
Once again, we can see that the Green's function as an inverse operator. Analogous to the previous subsection, the correlation function becomes,
\begin{eqnarray*}
\langle a(q)^\dagger a(q')\rangle&=&-2\int_{-\infty}^{\infty}d\omega\frac{\text{Im}G_{qq'}}{e^{\beta\omega}-1}\\
&=&-2\int_{-\infty}^{\infty}d\omega\frac{\text{Im}\frac{\delta_{qq'}}{2\pi(\omega-\hbar\omega_q)}}{e^{\beta\omega}-1}\\
&=&2\pi\int_{-\infty}^{\infty}d\omega \delta(\omega-\hbar\omega_q)\frac{\delta_{qq'}}{2\pi}\frac{1}{e^{\beta\omega}-1}\quad \quad \text{(using Plemjl formula (See Appendix \ref{s3}))}\\
\therefore \langle a(q)^\dagger a(q')\rangle &=& \frac{\delta_{qq'}}{e^{\beta\hbar\omega_q}-1}
\end{eqnarray*}
Again similar to the previous subsection, $a(q)^\dagger a(q)=n_q$ is the number operator and $\langle n_q\rangle$. The distribution function is given by,
\begin{eqnarray*}
\langle n_q\rangle=\frac{1}{e^{\beta\hbar\omega_q}-1}
\end{eqnarray*}

\section{Acknowledgment}
SS prepared these notes under the guidance of VS as part of a semester project in 2018 at UM-DAE CEBS. VS acknowledges support from the Raja Ramanna Fellowship by the DAE. SS acknowledges support from the INSPIRE scholarship for funding her undergraduate studies at UM-DAE CEBS. 
%Appendix
\section{Appendix}
\appendix
\renewcommand{\thesubsection}{\Alph{subsection}}

%commutation relations
\subsection{Commutation Relations$\colon$}\label{s4}
\begin{eqnarray}
[c_{i\sigma},c_{j\sigma'}^\dagger]_+&=&\delta_{ij}\delta_{\sigma\sigma'}\nonumber\\
\quad [c_{i\sigma},c_{j\sigma'}]_+&=& [c_{i\sigma}^\dagger,c_{j\sigma'}^\dagger]_+=0\Longrightarrow c_{i\sigma}^2=c_{j\sigma}^{\dagger 2}=0\nonumber\\
\quad [A,BC]_-&=&[A,B]_+C-B[A,C]_+\label{a2}\\
\quad [A,BC]_-&=&[A,B]_-C+B[A,C]_-\label{a3}\\
\quad [c_{i\sigma},\sum_{j\sigma'}n_{j\sigma'}]_-&=&\sum_{j\sigma'}c_{j\sigma'}\delta_{ij}\delta_{\sigma\sigma'}=c_{i\sigma}\label{a1}
\end{eqnarray}

%theta function
\subsection{$\theta$ Function}\label{s2}
\begin{eqnarray*}
\theta(t)&=&\int^t_{-\infty}e^{-\epsilon t}\delta(t)dt, \quad\quad \epsilon\rightarrow 0\quad(\epsilon>0)\\
\delta (t)&=&\int^{\infty}_{-\infty}e^{-ixt}dx\\
\text{then, } \theta(t)&=&\int^{\infty}_{-\infty}dx\int^t_{-\infty}e^{-(\epsilon+ix)t}dt\\
&=&\int^{\infty}_{-\infty}dx\frac{e^{-(\epsilon+ix)t}}{\epsilon+ix}\Big|_{-\infty}^t\\
&=&-\frac{1}{2\pi}\int^{\infty}_{-\infty}dx\frac{e^{-\epsilon t}e^{-ixt}}{\epsilon+ix}\\
&=&-\frac{1}{2\pi}\int^{\infty}_{-\infty}dx\frac{e^{-ixt}}{\epsilon+ix} \quad\quad e^{-\epsilon t}\rightarrow 1\text{ as }\epsilon\rightarrow 0\\
&=&\frac{i}{2\pi}\int^{\infty}_{-\infty}dx\frac{e^{-ixt}}{x+i\epsilon}
\end{eqnarray*}

%Plemjl's Formula
\subsection{Plemjl's Formula}\label{s3}
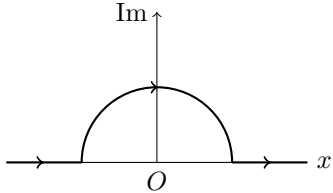
\begin{figure}[h]
\centering
\input{tikz41}
\caption{ Contour Diagram }
\label{b10}
\end{figure}

\begin{eqnarray*}
\text{Lim}_{\epsilon\rightarrow 0^+}\dfrac{1}{x\pm i\epsilon}=Pr\Big(\dfrac{1}{x}\Big)\mp i\pi\delta(x)
\end{eqnarray*}
This implies, 
\begin{eqnarray*}
\text{Lim}_{\epsilon\rightarrow 0^+}\int^\infty_{-\infty}\dfrac{f(x)}{x\pm i\epsilon}dx=Pr\int^\infty_{-\infty}\Big(\dfrac{f(x)}{x}\Big)dx\mp i\pi f(0)
\end{eqnarray*}
where f(x) is non singular in the neighbourhood of origin and $f(x)\rightarrow$ 0 as $x\rightarrow\pm\infty$. Now,
\begin{eqnarray*}
Pr\int^\infty_{-\infty}\Big(\dfrac{f(x)}{x}\Big)dx=\text{lim}_{\delta\rightarrow 0}\int^{-\delta}_{-\infty}\Big(\dfrac{f(x)}{x}\Big)dx+\int^{\infty}_{\delta}\Big(\dfrac{f(x)}{x}\Big)dx
\end{eqnarray*}
We take the help of Fig.(\ref{b10}) to prove the above mentioned formula.
 \\\textbf{R.H.S.}\\
Let C be contour (-$\infty$ to -$\delta$)+C$_{\delta}$+(+$\infty$ to +$\delta$). Thus,
\begin{eqnarray*}
 \int_C \dfrac{f(x)}{x}dx=Pr\int^\infty_{-\infty}\dfrac{f(x)}{x}+\int_{C_\delta}f(x)dx,
\end{eqnarray*}
where points on contour are taken as $\delta e^{i\theta}$ and near $\delta\rightarrow 0$, f(x)$\rightarrow$f(0). This would give us the R.H.S., 
\begin{eqnarray*}
\int_C \dfrac{f(x)}{x}dx=Pr\int^\infty_{-\infty}\dfrac{f(x)}{x}-i\pi f(0)
\end{eqnarray*}\\
\\\textbf{L.H.S.}\\
Let $C\rightarrow C'$, where $\infty\rightarrow \infty +i\epsilon$ and $-\infty\rightarrow -\infty +i\epsilon$. Then,
\begin{eqnarray*}
 \int_{C'}\dfrac{f(x)}{x}dx=\int_{-\infty+i\epsilon}^{\infty+i\epsilon}\dfrac{f(x)}{x}dx=\int_{-\infty}^{\infty}\dfrac{f(y+i\epsilon)}{y+i\epsilon}dy
\end{eqnarray*}
We assume $\epsilon$ to be infinitesimal, so $f(y+i\epsilon)\approx f(y)$. Therefore,
\begin{eqnarray*}
\int_{-\infty}^{\infty}\dfrac{f(y+i\epsilon)}{y+i\epsilon}dy=\int_{-\infty}^{\infty}\dfrac{f(y)}{y+i\epsilon}dy,\text{  or }\int_{-\infty}^{\infty}\dfrac{f(x)}{x+i\epsilon}dx
\end{eqnarray*}
On equating R.H.S and L.H.S we get the Plemjl's formula as follows.
\begin{eqnarray*}
\text{Lim}_{\epsilon\rightarrow 0^+}\int^\infty_{-\infty}\dfrac{f(x)}{x+ i\epsilon}dx=Pr\int^\infty_{-\infty}\Big(\dfrac{1}{x}\Big)dx- i\pi f(0)
\end{eqnarray*}

%Refrences

\end{document}

%% file: tikz11.tex
\begin{tikzpicture}
\draw (0,0) node[below]{$O$};
\draw (0,1) node{$\bullet$};
\draw (0,1) node[left]{$1$};
\draw (0,2) node[left]{$\theta$(t)};
\draw (2,0) node[right]{$t$};
\draw[->](0,0)--(0,2);
\draw[<-](-2,0)--(0,0);
\draw[<-](2,0)--(0,0);
\draw [dashed,ultra thick] (0,1) -- (2,1);
\draw [dashed,ultra thick] (0,0) -- (-2,0);
\end{tikzpicture}

%% file: tikz21.tex
\begin{tikzpicture}
\path [fill=gray!30] (0,0) rectangle (4.1,4.1) ;
\draw (1,0.9) node[below]{$k,\sigma$};
\draw (1,0.5) node[below]{(t)};
\node (1) [circle, draw, fill=black] at (1,1){};
\draw (3,3) node[above]{($t'$)};
\draw (3,3.4) node[above]{$k',\sigma'$};
\node (1) [circle, draw, fill=white] at (3,3){};
\draw [thick, decorate,decoration=snake] (1,1) -- (1,1.7);
\draw [thick,decorate,decoration=snake] (1,1.7) -- (1.7,2.3);
\draw [ultra thick][->,decorate,decoration=snake] (1.7,2.3) -- (1.7,2.5);
\draw [thick, decorate,decoration=snake] (1.7,2.5) -- (2.8,3);
\draw  (0,0) -- (0,4.1);
\draw  (0,4.1) -- (4.1,4.1);
\draw  (0,0) -- (4.1,0);
\draw (4.1,4.1) -- (4.1,0);

\end{tikzpicture}

%% file: tikz31.tex
\begin{tikzpicture}

  \begin{axis}[domain=-3.14:3.14, axis lines=none]
    \addplot [black, no markers]{1-cos(deg(x))-1}; 
\addplot [black, no markers,->] coordinates {(0,0) (4,0)};
\addplot [black, no markers,->] coordinates {(0,0) (-4,0)};
\addplot [black, no markers,->] coordinates {(0,0) (0,1.5)};
\addplot [black, no markers,->] coordinates {(0,0) (0,-1.5)};
\node at (axis cs:4,0) [anchor=west] {k};
\node at (axis cs:0,1.5) [anchor=south] {$\epsilon_k$};
\node at (axis cs:0,1) [anchor=west] {$(\mu+2t)$};
\node at (axis cs:0,-1) [anchor=north west] {$(\mu-2t)$};
\node at (axis cs:0,0) [anchor=south west] {$\mu$};
\node at (axis cs:3.14,0) [anchor=north] {$\pi/a$};
\node at (axis cs:-3.14,0) [anchor=north] {$-\pi/a$};
\node at (axis cs:0,0) {$\bullet$};
\node at (axis cs:0,1) {$\bullet$};
\node at (axis cs:0,-1) {$\bullet$};
\node at (axis cs:3.14,0) {$\bullet$};
\node at (axis cs:-3.14,0) {$\bullet$};
\end{axis}
\end{tikzpicture}

%% file: tikz41.tex
\begin{tikzpicture}
\draw (0,0) node[below]{$O$};
\draw (0,2) node[left]{Im};
\draw (2,0) node[right]{$x$};
\draw[thick] (1,0) arc (0:180:1 cm);
\draw[thick](1,0)--(2,0);
\draw[thick](-1,0)--(-2,0);
\draw[thick, <-](0.01,1)--(0.009,1);
\draw[thick, <-](1.51,0)--(1.509,0);
\draw[thick, ->](-1.51,0)--(-1.509,0);
\draw[->](0,0)--(0,2);
\draw(-2,0)--(0,0);
\draw(2,0)--(0,0);
\end{tikzpicture}